\documentclass[twocolumn,floats,showpacs,superscriptaddress,pre]{revtex4}

\usepackage{amsmath,amssymb,bm}
\usepackage{graphics}
\usepackage{epsfig}
\usepackage{graphicx}

\begin{document}

\title{Percolation on correlated networks}
\author{A. V. Goltsev}
\affiliation{Departamento de F{\'\i}sica da Universidade de Aveiro, 3810-193 Aveiro,
Portugal}
\affiliation{A.F. Ioffe Physico-Technical Institute, 194021 St. Petersburg, Russia}
\author{S. N. Dorogovtsev}
\affiliation{Departamento de F{\'\i}sica da Universidade de Aveiro, 3810-193 Aveiro,
Portugal}
\affiliation{A.F. Ioffe Physico-Technical Institute, 194021 St. Petersburg, Russia}
\author{J. F. F. Mendes}
\affiliation{Departamento de F{\'\i}sica da Universidade de Aveiro, 3810-193 Aveiro,
Portugal}

\begin{abstract}
We reconsider the problem of percolation on an equilibrium random
network with degree--degree correlations between nearest-neighboring
vertices focusing on critical singularities at a percolation
threshold. We obtain criteria for degree--degree correlations to be
irrelevant for critical singularities.
We present examples of networks in
which assortative and disassortative mixing leads to unusual
percolation properties and new critical exponents.

\end{abstract}

\maketitle





\section{Introduction} \label{intro}

Real-world networks are correlated
\cite{ab01a,dm01c,pvv01,Newman:n03a,blm06}. Correlations between degrees
of vertices in a network essentially characterise its structure.
Various real-world networks are markedly different in respect of
degree-degree correlations
\cite{pvv01,vpv02,Maslov:ms02,Maslov:msz02,Newman:n02}. In
particular, social networks show assortative mixing, i.e., a
preference of high-degree vertices to be connected to other
high-degree vertices, while technological and biological networks
are mostly disassortative, i.e., their high-degree vertices tend to
be connected to low-degree ones \cite{Newman:n02}. However, even the
simplest correlated networks with only pair correlations between the
nearest-neighbor degrees are still poorly understood. Our aim is to
find when the critical singularities for correlated networks of this
kind coincide with those for well-studied uncorrelated networks and
when and how much they differ.

At the present time it is well established that the small-world
effect and heterogeneity influence the cooperative dynamics and
critical phenomena of models defined on the top of complex networks
\cite{Dorogovtsev:dgm07}. However, numerous studies were devoted
mostly to more simple, uncorrelated networks. In an uncorrelated
network with a heavy-tailed degree distribution, critical
singularities of a continuous phase transition are characterized by
model dependent critical exponents which differ from the standard
mean-field ones and the critical exponents of two and three dimensional lattices, see, for example, Refs.~\cite{g89,sa03,b06}. The critical behavior depends on an asymptotic
behavior of a degree distribution at large degrees. For percolation
on an uncorrelated complex network that was demonstrated in
\cite{cah02,cha03}. One should expect, however, that for dynamical
processes taking place in a complex networks, correlations are
important. The simplest particular kind of correlations in a
networks are correlations between degrees of two nearest neighbors
in a network---so called degree--degree correlations. In this work
we consider only these specific, though representative,
correlations. Investigations of percolation
\cite{Newman:n02,Vazquez:vm03} and epidemic spreading
\cite{Boguna:bp-s02,Boguna:bpv03} demonstrated that the
degree--degree correlations strongly influence these phenomena. The
birth and growth of the giant connected component significantly
depends on the type of correlations---whether the degree--degree
correlations are assortative or disassortative. Compared to an
uncorrelated network with the same degree distribution, the
assortative correlations increase the resilience of a network
against random damage, while the disassortative correlations
diminish this resilience.

\begin{figure}[t]
\begin{center}
\scalebox{0.25}{\includegraphics[angle=0]{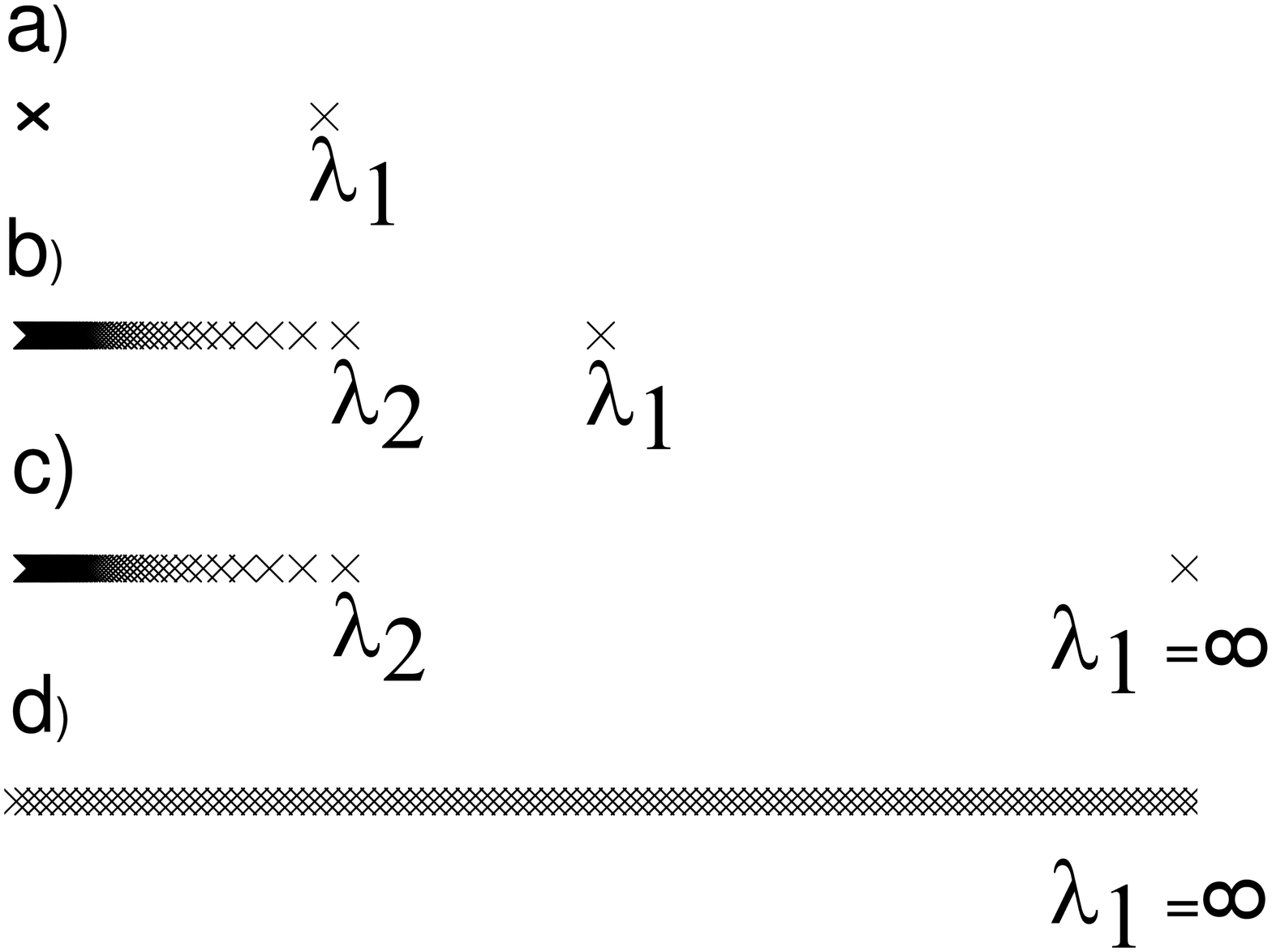}}
\end{center}
\caption{Schematic representation of eigenvalues (crosses) of the
branching matrix, Eq.~(\ref{B}), in the limit $N \rightarrow
\infty$. 
There are four situations:
(a) the largest eigenvalue
$\lambda_1$ is nonzero while all other eigenvalues are zero,
$\lambda_i=0$ for $i\geq 2$; (b) both the largest eigenvalue
$\lambda_1$ and the second largest eigenvalue $\lambda_2$ are
finite; (c) $\lambda_1$ diverges but $\lambda_2$ is finite; (d) the
sequence of eigenvalues is unbounded. Case (a) takes place
for uncorrelated networks but also can occur in specific disassortative networks. Cases (b)-(d) occur only in correlated networks.
} \label{fig0}
\end{figure}


\begin{table*}[t]
\caption{ Critical behavior of the order parameter $y\propto
(p-p_c)^\beta$, where $p$ is an occupation probability, in
correlated networks with a degree distribution $P(q)\sim q^{-\gamma
}$.
$\lambda_1$ and $ \lambda_2$ are the largest and second largest
eigenvalues of the branching matrix, Eq.~(\ref{B}). ``Weakly''
correlated networks, both assortative and disassortative, have the
same critical behavior as uncorrelated network with the identical
degree distribution, e.i., correlations are irrelevant. Model
(\ref{model}) with assortative mixing has a new critical exponent $\beta$ in the range $-1 <\alpha \leq 0$. Here, the
parameter $\alpha$ determines the average degree of the nearest
neighbor of a vertex with $q$ connections: $\overline{q}_{nn}(q)=
\text{const}+ q^{-\alpha}$. At $\alpha > 0$ this model is ``weakly''
correlated. Model (\ref{2-m2}) with disassortative mixing has a new critical
behavior at $\alpha > 0$. For this model, $\overline{q}_{nn}(q)= 1 +
q^{-\alpha}$. The $(\gamma,\alpha)$ phase diagrams of models
(\ref{model}) and (\ref{2-m2}) are shown in Figs.~\ref{phase1} and
\ref{phase2}, respectively. 
}
    \begin{tabular}{cccccc}
\hline
&&&&
\\[-11pt]
$\text{Network}$ & $\gamma$ &
Region on
& $\lambda$
&$p_c=1/\lambda_1$ & $\beta$
\\
&&
$(\gamma,\alpha)$ plane
&&&
\\[3pt]
\hline
&&&&
\\[-10pt]
uncorrelated or
& $ \gamma>4$&  &$\lambda_1, \lambda_2
<\infty$ & $ > 0$ & 1
\\
``weakly'' correlated
&  $3 <\gamma <4$ & &$\lambda_1, \lambda_2 <\infty$ &  $ > 0$ &
$1/(\gamma -3)$

\\

& $2<\gamma <3$ & & $\lambda_1 =\infty, \lambda_2 < \infty$&  0 &
$1/(3-\gamma)$

\\[3pt]
\hline
&&&&
\\[-10pt]
$\text{strongly assortative}$&  $\gamma >3 +\alpha$ & I &
$\lambda_1, \lambda_2 = \infty$& 0 & $(2-\gamma+\alpha)/\alpha$
\\

& $2<\gamma<3 +\alpha$ & II & $\lambda_1, \lambda_2 = \infty$& 0 &
$1/(3-\gamma )$
\\[3pt]
\hline
&&&&
\\[-10pt]

$\text{strongly disassortative}$ & $ \gamma>4-3 \alpha$ & I  &
$\lambda_1 <\infty, \,\,\, \lambda_2=0$ & $> 0$ & 1
\\

& $ 3-2\alpha < \gamma <4-3 \alpha$ & II & $\lambda_1 <\infty,
\,\,\, \lambda_2=0$ & $> 0$ & $(1-\alpha)/(\gamma - 3 +2\alpha)$

\\

& $ 2 < \gamma <3-2\alpha$ & III & $\lambda_1 =\infty, \,\,\,
\lambda_2=0$ & 0 & $(1-\alpha)/(3-\gamma - 2\alpha)$
\\[3pt]
\hline

\end{tabular}
\label{table1}
\end{table*}



One can construct an equilibrium network, where only degree--degree
correlations are present---the maximally random network with given
degree--degree correlations. This is impossible for non-equilibrium,
in particular, growing networks. Growing networks necessarily
demonstrate a wide spectrum of correlations and not only pair
correlations between degrees of the nearest neighbors. This type of
heterogeneity may result in an anomalous critical effect at the
birth point of a giant connected component
\cite{Callaway:chk01,Dorogovtsev:dms01b,kr02,l02,kkkr02,bb03,cb03,kd04}.
This transition resembles the Berezinskii-Kosterlitz-Thouless phase
transition in condensed matter
\cite{Berezinskii:b70,Kosterlitz:kt73}. The transition is explained
by a specific large-scale inhomogeneity of non-equilibrium networks
and related correlations.
The large-scale inhomogeneity here means the
difference in properties of vertices according to age
\cite{kr01,bp-s05}.

One should note that degree--degree correlations may also arise in
an equilibrium network if self-loops and multiple connections are
forbidden. In particular, this was demonstrated in
Ref.~\cite{lgkk06} for the static model of scale-free networks
\cite{Goh:gkk01}. Noh \cite{Noh:n07} observed an unusual critical
behavior in an exponential random graph model with a tunable
degree--degree correlations.

The earlier investigations
\cite{Newman:n02,Vazquez:vm03,Boguna:bp-s02,Boguna:bpv03} mostly
focused on the effect of degree--degree correlations on the
percolation and epidemic thresholds. In the present paper we
investigate the effect of degree--degree correlations in equilibrium
networks
on critical singularities at the percolation threshold.
We demonstrate that the critical behavior is determined by the
spectrum of a so-called \emph{branching matrix} and a degree
distribution. The eigenvalues of this matrix are real and can be
ordered in descending order. The largest eigenvalue determines the
percolation threshold and can be both finite and infinite depending
on a degree distribution in agreement with
Refs.~\cite{Newman:n02,Vazquez:vm03}. We derive necessary and
sufficient conditions for degree--degree correlations to be
irrelevant for critical singularities.
We give examples of strongly correlated networks with assortative
and disassortative mixing in which at least one of these conditions
is not fulfilled. These networks demonstrate new critical
singularities, see Table \ref{table1}. In particular, we propose
analytically treatable assortative networks with an unbounded
sequence of eigenvalues in the infinite size limit and show that
this peculiarity of the spectrum leads to a new critical behavior.
Remarkably, this network may be robust against a random damage even
if the second moment of its degree distribution is finite. This is
in contrast to uncorrelated networks which are robust only if the
second moment of a degree distribution diverges. We also study
specific disassortative networks which demonstrate new critical
singularities and are fragile. Remarkably, this takes place despite
the second moment of their degree distribution may be divergent and
the largest eigenvalue is finite. Our results are summed up in Table
\ref{table1} and Figs.~\ref{phase1} and \ref{phase2} which display
phase diagrams of two model networks with strong assortative and
disassortative mixing.

The paper is organized as follows. In Sec.~\ref{percolation} we give
a general description of correlations in a network and introduce the
branching matrix. In Sec.~\ref{ordinary percolation} we introduce a
basic equation describing percolation on a degree--degree correlated
network and reconsider the effect of assortative and disassortative
mixing on the percolation threshold.
In Sec.~\ref{critical} we find criteria for degree--degree
correlations to be irrelevant for critical singularities.
Sec.~\ref{model network} introduces a simple model of degree--degree
correlated networks which have a modular structure and an unbounded
sequence of eigenvalues of the branching matrix. In Sec.~\ref{new
class} we demonstrate that these networks have new critical
singularities. Disassortative networks with unusual critical
properties are studied in Sec.~\ref{disassort}. A detailed analysis
of the spectrum, its relationships with structural coefficients of a
correlated network and calculation of the entropy of the model
network are presented in Appendices \ref{spectrum}, \ref{network
coefficient} and \ref{entropy}, respectively.

\section{Degree--degree correlations}
\label{percolation}

We consider a random locally tree-like network of $N$ vertices in
which only pair correlations between nearest neighbor degrees $q$
and $q'$ are present.
We assume that there are vertices with degrees $q=q_{1},q_{2},
...q_{cut}$, and in total there are $N_d$ different values of degrees.
This correlated network is completely described by a given symmetric
joint degree-degree distribution, $P(q,q')=P(q',q)$, otherwise the
network is homogeneously random. The degree distribution $P(q)$ can
be calculated from $P(q,q')$ as follows:
\begin{equation}
\sum_{q'}P(q',q)=qP(q)/\langle q \rangle ,  \label{dd}
\end{equation}
where the brackets $\langle...\rangle$ denote an average over the
degree distribution $P(q)$, for example, $\langle q \rangle =
\sum_{q}qP(q)$. Below for brevity we use the notations: $z_1 \equiv
\langle q \rangle $ and $ z_2\equiv \langle q(q-1)\rangle$. It is
convenient to introduce the conditional probability,
\begin{equation}
P(q'|q)=\frac{z_{1} P(q,q')}{qP(q)},  \label{cp}
\end{equation}
that if an end vertex of an edge has degree $q$, then the second end
has degree $q'$. The functions $P(q,q')$ and $P(q'|q)$ are
normalized:
\begin{equation}
\sum_{q,q'}P(q,q')=\sum_{q'}P(q'|q)=1.
\label{norma}
\end{equation}
In uncorrelated networks the conditional probability does not depend
on $q$: $P(q'|q) = q'P(q')/z_{1}$.
We define a
non-symmetric \emph{branching matrix} as
follows:
\begin{equation}
B_{qq'}=(q'-1)P(q'|q). \label{B}
\end{equation}
This matrix has the following property:
\begin{equation}
q(q-1)P(q)B_{qq'}=q'(q'-1)P(q')B_{q'q}. \label{B-t}
\end{equation}
According to definition (\ref{B}), an entry $B_{qq'}$ of this matrix
is equal to the branching coefficient $q'-1$ of an edge, which
emanates from a vertex with degree $q$ and has a vertex with degree
$q'$ at the second end, multiplied by the probability $P(q'|q)$ that
the second end has degree $q'$. In Secs.~\ref{ordinary percolation}
and \ref{critical} we will show that the structure of the spectrum
of this matrix determines the critical properties of the
percolation. Relationship of this spectrum with clustering
coefficients and the mean intervertex distance in a correlated
network is considered in Appendix~\ref{network coefficient}.

Using the branching matrix we can calculate the average branching
coefficient $B(q)$ of an edge which emanates from a vertex of degree
$q$:
\begin{equation}
B(q)=\sum_{q'} B_{qq'}.
\label{Bq}
\end{equation}
The coefficient $B(q)$ is related
to the average nearest-neighbor's degree of the vertices of degree $q$:
\begin{equation}
\overline{q}_{nn}(q)=\sum_{q'} q'P(q'|q)=B(q)+1. \label{annd}
\end{equation}
A mean branching coefficient $\overline{B}$ is equal to
\begin{equation}
\overline{B}=\frac{1}{z_1} \sum_q
B(q)qP(q)=\frac{z_2}{z_1}=\frac{\langle q^2 \rangle}{\langle q
\rangle}-1.
\label{meanB}
\end{equation}
It only depends on the degree distribution $P(q)$. An integral characteristic of degree-degree correlations is given by
the Pearson coefficient \cite{Newman:n02}:
\begin{equation}
r= \frac{1}{z_1 \sigma^2}\sum_{q}[B(q)-\overline{B}]q(q-1)P(q),
 \label{r}
\end{equation}
where
$$
\sigma^2 {=} \sum_q (q{-}1)^2 \,\frac{qP(q)}{z_1}\, - \left[\sum_q (q{-}1)\frac{qP(q)}{z_1}\right]^2
\!\!\!\!=\! \frac{\langle q^3 \rangle}{\langle q \rangle} {-} \frac{\langle q^2 \rangle^2}{\langle q \rangle^2}
$$
is for normalization. The Pierson coefficient
is positive for assortative mixing and negative for disassortative
mixing.

In an uncorrelated complex network, $\overline{q}_{nn}(q)$ does not
depend on $q$: $\overline{q}_{nn}(q)=\overline{B}+1$. In contrast,
in networks with assortative mixing, $\overline{q}_{nn}(q)$
increases with increasing $q$ while in disassortative networks it
decreases. For example, $\overline{q}_{nn}(q)$ shows a power law
decay $\overline{q}_{nn}(q) \propto q^{- 0.5}$ for the Internet
\cite{pvv01,vpv02}.
Recursive scale-free networks, growing by the linear
preferential attachment mechanism, have growing asymptotics $\overline{q}_{nn}(q)\sim \ln q$ at large $q$
at $\gamma >3$ and
decaying $\overline{q}_{nn}(q)$ as a power law at
$2< \gamma <3$
\cite{bp-s05,kr02}.

\section{Tree ansatz equations and percolation threshold}
\label{ordinary percolation}

A key quantity in the percolation problem is the probability $x_q$
that if an edge is attached to a vertex of degree $q$, then,
following this edge to its second end, we will not appear in a giant
connected component \cite{Vazquez:vm03}. The number of
unknown order parameter components $x_q$ is equal to the number of different
degrees, $N_d$. In contrast, only a one-component order parameter $x$ describes
percolation on an uncorrelated complex network \cite{Callaway:cns00,nsw01}.

Let $p$ be the probability that a vertex is retained in a randomly
damaged network.
Within a tree-ansatz theory, which assumes that a network has a
locally tree-like structure, equations for the probabilities $x_q$ and
the relative size $S$ of a giant connected component have the
following form \cite{Vazquez:vm03}: 
\begin{eqnarray}
&& x_q = 1-p+p\sum_{q'}P(q'|q) (x_{q'})^{q'-1} , \label{xq}
\\[5pt]
&& 1-S = 1-p+p\sum_{q} P(q) (x_q)^{q}.
\label{s}
\end{eqnarray}
Equations~(\ref{xq}) and (\ref{s})
directly generalize equations derived for
percolation on an uncorrelated complex network
\cite{Callaway:cns00,nsw01}, where $x_q=x$, see also
Refs.~\cite{dm01c,Newman:n03a,Dorogovtsev:dgm07}.
The set of
Eqs.~(\ref{xq}) determines unknown probabilities $x_q$ for
$q=q_1, q_2,...q_{cut}$. Newman \cite{Newman:n02} originally derived
these equations using generating functions, and numerically solved
them for various networks. The analysis of
Eqs.~(\ref{xq}) shows that the birth of the giant connected component is
a continuous phase transition. Below the percolation threshold,
i.e., at $p<p_c$, these equations only have a trivial solution
$x_{q} =1$, and there is no giant connected component. A giant connected component is present
above the percolation point, $p>p_c$, where there is a solution with
$x_{q} < 1$. Introducing a parameter $y_q=1-x_q$, we rewrite
Eq.~(\ref{xq}) as follows:
\begin{equation}
p \sum_{q'} B_{qq'}y_{q'}-y_q=p \sum_{q'=3}^{q_{cut}}
\sum_{n=2}^{q'-1}{q'-1 \choose n}(-y_{q'})^n P(q'|q). \label{yq}
\end{equation}
One can solve this set of equations near the critical point,
$\tau=p/p_{c} -1\ll 1$ when $y_{q} \ll 1$.

First we study the percolation threshold for a
degree--degree correlated network. We will use the spectral
properties of the branching matrix, Eq.~(\ref{B}), so let us
remind some basics. Eigenvalues $\lambda_i$ and eigenvectors
$\Phi^{(i)}_{q}$ associated with these eigenvalues are defined by
the equation:
\begin{equation}
\sum_{q'} B_{qq'}\Phi_{q'}^{(i)}= \lambda_i \Phi_q^{(i)},
\label{sp}
\end{equation}
where the index $i$ labels the eigenvalues, $i=1,2,\ldots, N_d$. We
consider the case of positive entries: $B_{qq'} >0$. The following
statements hold. (i) The largest eigenvalue $\lambda_1$ is positive.
(ii) The entries $\Phi_{q}^{(1)}$ of the maximal eigenvector are
positive, $\Phi_{q}^{(1)} >0$. (iii) All eigenvalues $\lambda_i$ are real and can be
ordered:
$\lambda_1 > \lambda_2 > ...
> \lambda_{N_d}$. (iv) The eigenvectors $\Phi^{(i)}$ associated with
these eigenvalues form a complete orthonormal basis set. That is,
\begin{eqnarray}
&&\sum_{q} w(q) \Phi_{q}^{(i)} \Phi_{q}^{(j)}= \delta_{i,j},
\label{n1}
\\[5pt]
&& w(q)\sum_{\lambda} \Phi_{q}^{(\lambda)} \Phi_{q'}^{(\lambda)}=
\delta_{q,q'}, \label{n2}
\end{eqnarray}
where $w(q)=q(q-1)P(q)/z_1$ is a weight function, $\delta_{i,j}$
and $\delta_{q,q'}$ are the Kronecker symbols. Properties (i) and
(ii) follows from the Perron-Frobenius theorem, see, for example,
Ref. \cite{Minc}. Properties (iii) and (iv) are proved in
Appendix~\ref{spectrum}. One can show that for an uncorrelated network there is a
single largest eigenvalue and an ($N_d -1$)-degenerate zero eigenvalue:
\begin{equation}
\lambda_1=z_{2}/z_1, \,\,\,\, \lambda_{i\geq 2}=0.
\label{unc-spectrum}
\end{equation}
The percolation threshold corresponds to a critical probability
$p_c$ above which a nontrivial solution of Eq.~(\ref{yq}), $y_q \geq
0$, appears. Taking into account only the linear terms, we get the
following condition:
\begin{equation}
y_q=p \sum_{q'} B_{qq'}y_{q'}. \label{yq-linear}
\end{equation}
We represent $y_q$ as a linear combination of the mutually
orthogonal eigenvectors which we call \emph{modes}:
\begin{equation}
y_q = \sum_{i=1}^{N_d} a_i \Phi_{q}^{(i)}. \label{expansion}
\end{equation}
The amplitudes $a_i$ are unknown functions of $p$. It is obvious
that $a_i=0$ at $p <p_c$. Substituting Eq.~(\ref{expansion}) into
Eq.~(\ref{yq-linear}) and using the orthogonality Eq.~(\ref{n1}) of
the eigenvectors, we obtain an equation $a_i=p\lambda_i a_i$. One
can see that a nontrivial solution $a_1 \neq 0$ appears when
$p\lambda_1=1$.
So the mode associated with $\lambda_1$ is critical.
This gives the following criterion for
the percolation threshold found in Refs.~\cite{Newman:n02,Vazquez:vm03}:
\begin{equation}
p_c \lambda_1 =1. \label{pc}
\end{equation}
Thus the generalization of the Molloy--Reed criterion to
undamaged correlated networks, i.e., at $p=1$, is the following
condition: {\em if the largest eigenvalue $\lambda_1$ of the
branching matrix, $B_{qq'}$, is larger than $1$, then the correlated
network has a giant connected component.}
In uncorrelated networks the criterion (\ref{pc}) is reduced to the
well-known one: $z_2/z_1 = 1$.

In Appendix \ref{spectrum} we prove that at a given degree
distribution $P(q)$, the largest eigenvalue $\lambda_{1}^{(as)}$ of
an assortative network is larger then $z_{2}/z_1$ while in a
disassortative network it is smaller:
\begin{equation}
\lambda_{1}^{(dis)} < \frac{z_2}{z_1}< \lambda_{1}^{(as)}.
 \label{3-lambda}
\end{equation}
There are the following lower and upper boundaries for $\lambda_1$
(see Appendix \ref{spectrum}):
\begin{equation}
\frac{1}{z_2} \sum_q B(q)q(q-1)P(q) \leq \lambda_1 \leq  \max_q
B(q).
 \label{4-lambda}
\end{equation}
Suppose that a degree distribution is such that $z_2 /z_1=1$.
Then inequality (\ref{3-lambda}) leads to the following statements.
(i) An uncorrelated network with this degree distribution is at the birth point of a giant connected component. (ii) An assortative network with this degree distribution has a giant connected component. (iii) A disassortative network with this degree distribution has no giant connected component.
According to criterion (\ref{pc}) and inequality (\ref{3-lambda}),
percolation thresholds in randomly damaged assortative and
disassortative networks with the same degree distribution satisfy the
inequality:
\begin{equation}
p_{c}^{(as)} < z_1 /z_2 < p_{c}^{(dis)}.
\label{pc-en}
\end{equation}
Therefore, assortative mixing enhances resilience of a correlated
network against random damage while a disassortative mixing
decreases it \cite{Newman:n02}.

If the largest eigenvalue $\lambda_1$ diverges in the infinite
network, then the percolation threshold $p_c$ tends to zero. In this
case the giant connected component of a correlated network cannot be
eliminated by a random removal of vertices \cite{Vazquez:vm03}.
In an assortative network, in accordance with inequality (\ref{3-lambda}), this takes place when the
second moment $z_2$ diverges. This criterion ($z_2\to\infty$) of the robustness of an
uncorrelated network was found in \cite{Albert:ajb00,cah02}. Remarkably, $\lambda_1$ can diverge even if an assortative
network has a finite $z_2$ but degree--degree correlations are
sufficiently strong. An example of an assortative network of this kind is given
in Sec.~\ref{model network}.
On the other hand, Va\'{a}zquez and Moreno \cite{Vazquez:vm03}
found a network with strong disassortative mixing, which has
a finite percolation threshold and is fragile against a
random damage even if the second moment $z_2$ diverges.

\section{Critical behavior of ``weakly'' correlated networks}
\label{critical}

In this section we derive necessary and sufficient conditions for
degree--degree correlations to be irrelevant for critical
singularities at the percolation transition.
These conditions are:
\begin{itemize}

\item[(I)] The largest eigenvalue $\lambda_1$ of the branching matrix must be finite if $\langle q^2
\rangle$ is finite, or $\lambda_1 \rightarrow \infty$ if $\langle q^2 \rangle \rightarrow \infty$.
\item[(II)]  The second largest eigenvalue $\lambda_2$  must be finite.
\item[(III)] A sequence of
entries $\Phi_{q}^{(1)}$ of the maximum eigenvector of this matrix
must converge to a finite non-zero value at $q \rightarrow
\infty$, see Eq.~(\ref{bounds}).

\end{itemize}

Let us solve Eq.~(\ref{yq}) near the percolation transition. We
use the fact that the eigenvectors $\Phi_{q}^{(i)}$ of the branching matrix
$\widehat{B}$ form the complete orthogonal basis set, see
Eqs.~(\ref{n1}) and (\ref{n2}). Substituting Eq.~(\ref{expansion})
into Eq.~(\ref{yq}), we get a set of nonlinear equations for unknown
amplitudes $a_i$:
\begin{equation}
(p \lambda_i -1)a_i =p \lambda_i S_i(\textbf{a}). \label{yq-m}
\end{equation}
Here $S_i(\textbf{a})$ is a function of the amplitudes
$\textbf{a}=(a_1, a_2, ...)$:
\begin{equation}
S_i(\textbf{a})=\frac{1}{z_1} \sum_{q=3}^{q_{cut}}
\sum_{n=2}^{q-1}{q-1 \choose n}(-y_{q})^n qP(q)\Phi_{q}^{(i)}.
\label{yq-S}
\end{equation}

Let us find $a_i$ in the leading order in $\tau \ll 1$ under
conditions I and II. Substituting Eq.~(\ref{expansion}) into
Eqs.~(\ref{yq-m}) and (\ref{yq-S}) and taking into account only the
quadratic terms in $a_i$, we get a set of approximate equations for
the amplitudes $a_i$:
\begin{equation}
(p\lambda_i - 1)a_i=p\lambda_i \sum_{m,n} M_{imn}a_m a_n,
 \label{1-mode}
\end{equation}
where
\begin{equation}
M_{imn}= \frac{1}{2 z_{1}} \sum_q q(q-1)(q-2)P(q)
\Phi_{q}^{(i)} \Phi_{q}^{(n)} \Phi_{q}^{(m)}.
 \label{Mimn}
\end{equation}
Let us first consider a degree--degree correlated network with
finite coefficients $M_{imn}$ in the infinite size limit $N
\rightarrow\infty$. In the leading order in $\tau$, equation
(\ref{1-mode}) has a solution
\begin{eqnarray}
&&a_1 \simeq \frac{\tau}{M_{111}} + O(\tau ^2),
\label{a1}
\\[5pt]
&& a_i \simeq - \frac{\lambda_i a_{1}^2 M_{i11}}{(\lambda_1
-\lambda_i)} \propto O(\tau ^2) \ \ \ \text{for $i\geq 2$}.
\label{a2}
\end{eqnarray}
One can see that the amplitude $a_1$ of the critical mode
$\lambda_1$ has the standard mean-field dependence. It is much
larger than the amplitudes $a_i$ of the modes $i\geq 2$ which we
call \emph{transverse modes}. Therefore
near the percolation threshold the order parameter
$y_q=1-x_q$ is mainly determined by the critical mode $\Phi_{q}
^{(1)}$ associated with the largest eigenvalue. The transverse modes
give a smaller contribution:
\begin{equation}
y_q \approx a_1 \Phi_{q}^{(1)} + O(a_{1}^2).
 \label{yq-approx}
\end{equation}
Above we assumed that the coefficient $M_{111}$ is finite. This
assumption takes place if the sequence of entries $\Phi_{q}^{(1)}$
converges at $q \rightarrow q_{cut}\rightarrow \infty$, e.i.,
$\lim_{q \rightarrow \infty} \Phi_{q}^{(1)}< \infty$. If this
limiting value is non-zero,
\begin{equation}
0< \lim_{q \rightarrow \infty} \Phi_{q}^{(1)} < \infty,
 \label{bounds}
\end{equation}
then the third moment $\langle q^3 \rangle$ of the degree
distribution must be finite, $\langle q^3 \rangle < \infty$.
Equation (\ref{bounds}) is the condition III formulated above.
Thus we conclude that under conditions I--III the
percolation in a correlated scale-free networks with $\gamma >4$ has
the standard mean-field critical behavior, Eq.~(\ref{a1}), with the
critical exponent $\beta=1$.
In Sec.~\ref{disassort} we will show that the case $\lim_{q
\rightarrow \infty} \Phi_{q}^{(1)}=0$, e.i., the condition III is broken down, corresponds to a strongly
correlated network.

Consider the case $3 < \gamma \leq 4$ under conditions
I--III. In this case it is necessary to take into
account all orders in $a_1$ in Eq.~(\ref{yq-m}). The reason is that
the coefficients of the expansion over $a_1$ diverge as $N
\rightarrow \infty$. It is this divergence that leads to the
non-standard critical behavior of percolation on an uncorrelated
complex network \cite{cah02,cha03}. We assume that the inequality
$a_1 \gg a_i$ for all $i\geq 2$ is also valid at $3 < \gamma \leq
4$. Below we will confirm this assumption. We expand the function
$S_i(\textbf{a})$ over the amplitudes $a_i$ of the transverse modes
and save only linear terms in $a_i$:
\begin{equation}
S_i(\textbf{a})\simeq s_{i}(a_1) + \sum_{j\geq 2} s_{ij}(a_1)a_j +
...  , \label{yq-S2}
\end{equation}
where
\begin{eqnarray}
&& s_{i}(a_1)\equiv S_i(a_1,0,0,\ldots),
\label{si1}
\\[5pt]
&& s_{ij}(a_1)\equiv \frac{\partial S_i(\textbf{a})}{\partial
a_j}|_{\textbf{a}=(a_1,0,0,..)}.
\label{sij1}
\end{eqnarray}
The function $s_{i}(x)$ is given by the following series:
\begin{equation}
s_i(x)=\sum_{n=2}^{^{q_{cut}-1}} K_{i}(n)(-x)^n,  \label{si-x}
\end{equation}
where
\begin{equation}
K_{i}(n)=\frac{1}{z_1} \sum_{q=n+1}^{q_{cut}}{q-1
\choose n}qP(q)(\Phi_{q}^{(1)})^{n}\Phi_{q}^{(i)}. \label{Bin}
\end{equation}
Similarly, the function $s_{ij}(x)$ is
\begin{eqnarray}
&& s_{ij}(x) = \sum_{n=2}^{^{q_{cut}-1}} K_{ij}(n)(-x)^n,
\label{sij-x}
\\[5pt]
&& \!\!\!\!\!\! \!\!\!\!\!\!\!\!\!\!\! K_{ij}(n)=\frac{n}{z_1} \! \sum_{q=n+1}^{q_{cut}}{q-1 \choose
n}qP(q)(\Phi_{q}^{(1)})^{n-1}\Phi_{q}^{(i)}\Phi_{q}^{(j)}.
\label{sij-x}
\end{eqnarray}
At $3 < \gamma \leq 4$ the functions $s_{i}(x)$ and $s_{ij}(x)$ are
singular functions of $x$ because the coefficients $K_{i}(n)$ and
$K_{ij}(n)$ diverge as $q_{cut}(N)\rightarrow\infty$. Using the
method of summation used in \cite{cah02,Goltsev:gdm03} we get
asymptotic results:
\begin{eqnarray}
&& s_i(x) \approx b_i x^{\gamma -2},  \nonumber
\\[5pt]
&& s_{ij}(x)\approx b_{ij} x^{\gamma -3}. \label{s-asymp}
\end{eqnarray}
Here the numerical coefficients $b_i$ and $b_{ij}$, where
$i,j=1,2,... \,$, depend on a specific model of a correlated
network. Substituting Eqs.~(\ref{yq-S}) and (\ref{s-asymp}) into
Eq.~(\ref{yq-m}) we obtain a set of nonlinear equations
\begin{equation}
(p \lambda_i -1)a_i =p \lambda_i [b_i a_{1}^{\gamma -2} +
a_{1}^{\gamma -3} \sum_{j \geq 2} b_{ij}  a_j ] \label{modes-2}
\end{equation}
for $i=1,2, ... \,$. They have an asymptotic solution
\begin{eqnarray}
&&a_1 \approx \Bigl(\frac{\tau}{b_{1}}\Bigr)^{1/(\gamma -3)}, \label{a1-asymp}
\\[5pt]
&& a_i \approx - \frac{b_i \lambda_i a_{1}^{\gamma - 2}}{(\lambda_1
-\lambda_i)} \propto O(\tau ^{(\gamma - 2)/(\gamma - 3)}).
\label{a2-asymp}
\end{eqnarray}
In the case $3 < \gamma \leq 4$ the amplitude $a_1$ of the critical
mode $\Phi_{q}^{(1)}$ also is mush larger than the amplitudes of the
transverse modes, i.e., $a_1 \gg a_i$ for all $i\geq2$. So, in the leading order, we get
$y_q \approx a_1 \Phi_{q}^{(1)}$. The critical
exponent $\beta$ is equal to $1/(\gamma -3)$ as for an uncorrelated network
with the identical degree distribution.

Now consider the case $2 < \gamma \leq 3$ under conditions (I)--(III). According to
Eq.~(\ref{3-lambda}), in an assortative network with a divergent
second moment $z_2$ the largest eigenvalue $\lambda_1$ diverges in
the limit $N \rightarrow \infty$. The percolation threshold is $p_c
=0$, and a giant connected component is present at any $p>0$. One
can show that at $p\ll 1$ the order parameter $y_q$ and the size of giant
connected component $S$ are
\begin{eqnarray}
&& y_q\propto p^\beta \Phi_{q}^{(1)}, \label{a3}
\\[5pt]
&& S \propto p^{1+\beta} \label{a4}
\end{eqnarray}
where $\beta=1/(3-\gamma)$. This is the same
singularity as was found in uncorrelated networks with $2 < \gamma <
3$ \cite{cah02,cha03}. This result is also valid for
disassortative networks with $\lambda_1=\infty$.

The main conclusion of this section is that under conditions
(I)--(III), a correlated complex network has the same
critical behavior as an uncorrelated random network with the
identical degree distribution.
Our results are summed up in Table~\ref{table1}.

Degree-degree correlations may be strong and violate one of these
conditions. In Sec.~\ref{new class} and Sec.~\ref{disassort} we will
give examples of such networks and new critical singularities.

\section{Exactly solvable model of a correlated network}
\label{model network}

In this section we introduce a simple model of a correlated network
which allows analytical treatment. This, actually, toy model permits
us to check general results of the percolation theory derived above
and to study an effect of strong assortative mixing. This model is
also interesting due to its clear modular structure.

Let us consider a correlated network with a given degree
distribution $P(q)$ and a degree-degree distribution $P(q,q')$ which
is factorized at $q\neq q'$ as follows:
\begin{eqnarray}
&&P(q,q')=\delta_{q,q'}[f(q)+1]\rho^{2}(q) P^{2}(q) \nonumber
\\[5pt]
&& +(1-\delta_{q,q'})\rho(q) P(q)\rho(q') P(q'). \label{model}
\end{eqnarray}
Compare this with an uncorrelated network where $P(q,q')=q P(q)q'
P(q')/z_{1}^2$ which corresponds to $\rho(q)=q/z_1$. The function
$f(q)$ is non-negative for assortative mixing. In this case vertices
at the ends of an edge have the same degree with a higher probability
rather than different degrees. In the limit $f(q) \rightarrow
\infty$ we have a set of disconnected random $q$-regular modules.
For disassortative mixing, we choose $-1 < f(q)\leq 0$, then the probability to
have different degrees at the ends of an edge is higher than to have the
same degree.

For a given degree distribution $P(q)$ and a function $f(q)$,
the substitution of Eq.~(\ref{model}) into Eq.~(\ref{dd}) gives the
following equation for the function $\rho(q)$:
\begin{equation}
\rho(q)=\frac{q}{z_1 \rho} \frac{2}{1+[1 +4qf(q)P(q)/(z_1
\rho)]^{1/2}},
 \label{rho-q}
\end{equation}
where the parameter $\rho$ must be found self-consistently from the
equation:
\begin{equation}
\rho= \sum_{q} \rho(q)P(q).
 \label{rho}
\end{equation}
For model (\ref{model}), eigenvalues $\lambda$ and eigenvectors
$\Phi^{(\lambda)}_q$ of the branching matrix, Eq.~(\ref{B}), are
determined by the following exact equations:
\begin{eqnarray}
&&1=z_1 \sum_{q} \frac{(q-1)}{q} \frac{\rho^{2}(q)P(q)}{\lambda -
\Lambda(q)},
 \label{lambda-model}
\\[5pt]
&& \Phi^{(\lambda)}_q=\frac{z_1 \rho(q) n(\lambda)}{q[\lambda -
\Lambda(q)]}.
 \label{eigenv-model}
\end{eqnarray}
Here we have introduced a function
\begin{equation}
\Lambda(q)=\frac{(q-1)}{q} z_1 f(q)\rho^{2}(q)P(q).
 \label{eigenf-model}
\end{equation}
$n(\lambda)$ is a normalization constant determined by
Eq.~(\ref{n1}). Choosing a function $f(q)$ and degree distribution
$P(q)$, we obtain a correlated network with a specific spectrum of
the eigenvalues of the branching matrix.

For model (\ref{model}) the average branching parameter $B(q)$
defined by Eq.~(\ref{Bq}) is equal to
\begin{equation}
B(q)=\frac{z_1 \rho(q)}{q} \sum_{q'} (q'-1)\rho(q')P(q') + \Lambda
(q).
 \label{z2q-m}
\end{equation}
Assuming that $P(q)f(q) \ll  1$ at large $q$, we
obtain the asymptotic behavior: $\rho(q)\approx q/(\rho z_1)$. Therefore, $B (q)= \text{const} + \Lambda (q)$ at $q \gg 1$.
If $\Lambda (q) \propto q^{-\alpha}$ and $-1
\leq \alpha < 0$, $B (q)$ increases with increasing $q$.

\begin{figure}[t]
\begin{center}
\scalebox{0.3}{\includegraphics[angle=0]{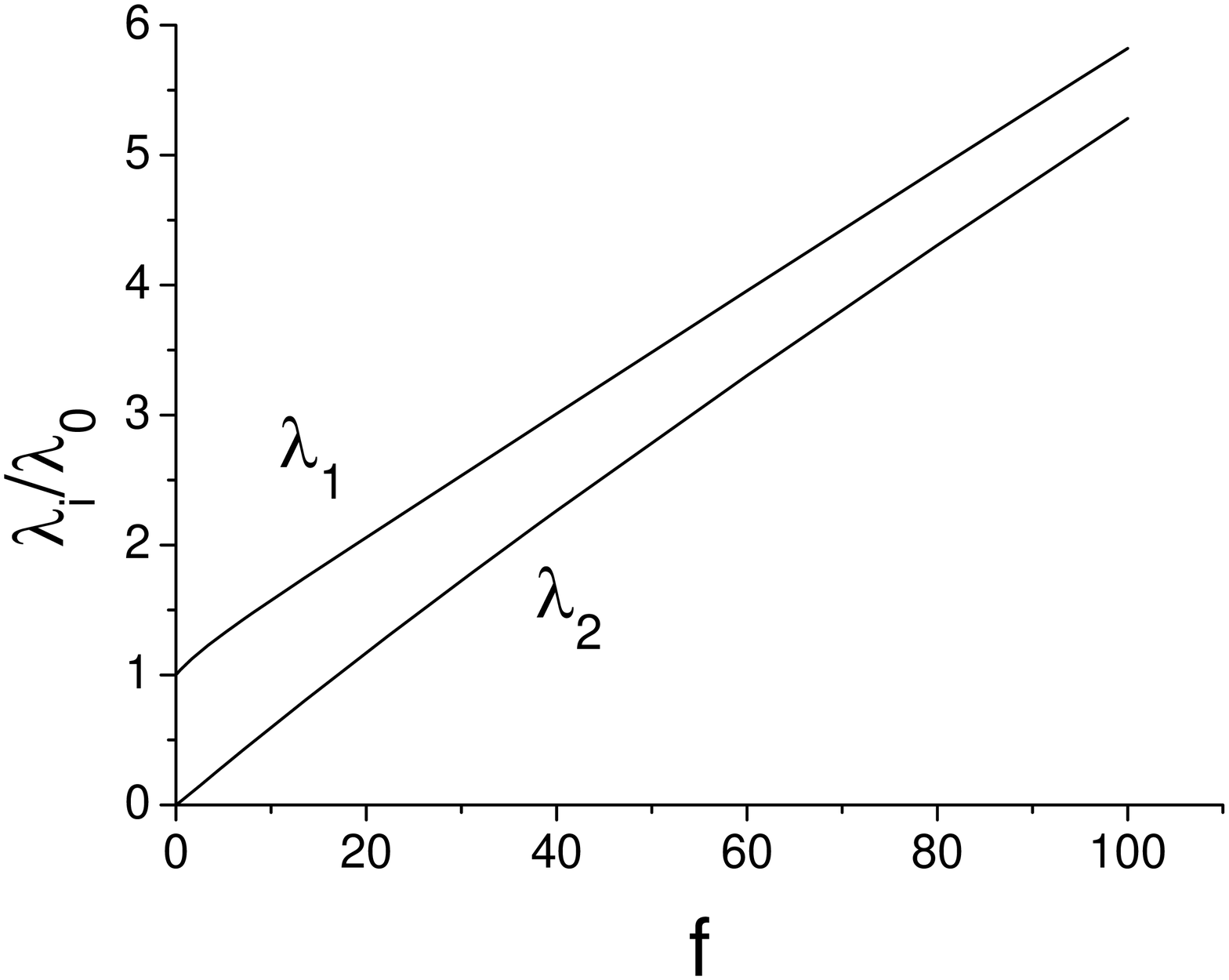}}
\end{center}
\caption{Dependence of the largest, $\lambda_1$,
and the second largest, $\lambda_2$, eigenvalues
of the branching matrix $\widehat{B}$ on the
parameter $f$ in the uniform case $f(q)=f$ for model
(\ref{model}) with a scale-free degree distribution $P(q)=Aq^{-4}$,
$\langle q\rangle =2.3$, $\langle q(q-1)\rangle =11.3$. Other
eigenvalues are non-negative and lie below $\lambda_2$. The
eigenvalues are normalized in respect to $\lambda_0 = z_2/z_1$.}
\label{eigen}
\end{figure}
\begin{figure}[t]
\begin{center}
\scalebox{0.3}{\includegraphics[angle=0]{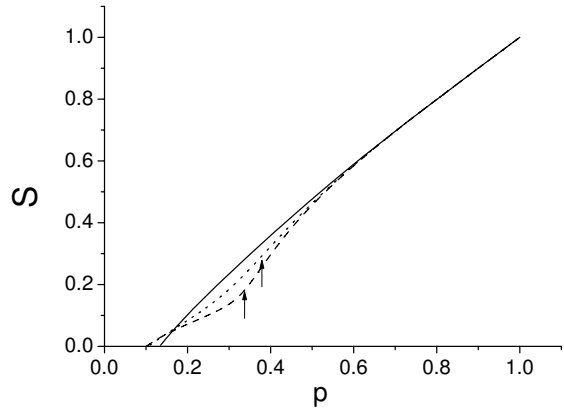}}
\end{center}
\caption{Relative size $S$ of a giant connected component versus
the occupation probability $p$ for correlated network
(\ref{model}) having two groups of vertices with degrees $q_1 =4$ and
$q_2 =11$, $P(q_1)=0.6$ and $P(q_2)=0.4$.
The curves were obtained by
numerical solution of Eq.~(\ref{xq}) in the following cases:
(i) $f(q_1)=f(q_2)=0$, i.e., an uncorrelated network (solid line);
(ii) an assortative network with $f(q_1)=f(q_2)=10$ (dotted line);
(iii) an assortative network with $f(q_1)=f(q_2)=100$ (dashed line).
Arrows indicate the positions of $1/\lambda_2$
for (ii) and
(iii).} \label{fig1}
\end{figure}
\begin{figure}
\begin{center}
\scalebox{0.3}{\includegraphics[angle=270]{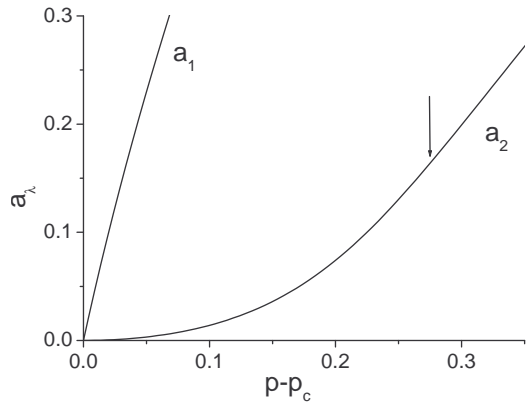}}
\end{center}
\caption{Amplitudes $a_1$ and
$a_2$ of critical and transverse modes, respectively,
versus $p-p_c$ in an assortative network with degree--degree
distribution (\ref{model}) and the same parameters as for the dotted curve in
Fig.~\ref{fig1}: $f(q_1)=f(q_2)=0$.
The percolation threshold is $p_c=1/\lambda_1$. The arrow shows the
point $p=1/\lambda_2$. } \label{fig3}
\end{figure}

Model (\ref{model}) has a modular structure. Modules are formed by
vertices with equal degrees $q$. The number of modules is equal to
the number of different degrees, $N_d$. In the limit
$f(q)\rightarrow \infty$ for all $q$, this network consists of
disconnected modules. Each module is a $q$-regular random graph.
In this limit, the Pearson coefficient $r=1$, which
corresponds to a perfectly assortative network
\cite{Newman:n02}. If $f(q)$ is large but finite we get a weakly
connected modules.

Equations~(\ref{lambda-model}) and (\ref{eigenv-model}) show that
the largest eigenvalue $\lambda_1$ is larger than  $\max \Lambda(q)$.
All other eigenvalues lie in the range: $\min \Lambda(q) < \lambda <
\max \Lambda(q)$. For example, if the function $\Lambda(q)$ is a monotonously increasing function, then there is only one
eigenvalue in every interval $(\Lambda(q_i),\Lambda(q_{i+1}))$.
If $\max \Lambda(q)< \infty$, then $\lambda_2$ is finite because
$\lambda_2 < \max \Lambda(q)$. $\lambda_1$ is finite at $\langle q^2
\rangle < \infty$ and diverges at $\langle q^2 \rangle \rightarrow
\infty$. Equation (\ref{eigenv-model}) shows that the
maximal eigenvector satisfies Eq.~(\ref{bounds}). Thus conditions
I--III in Sec. \ref{critical} are fulfilled if $\max \Lambda(q)$ is finite. In this
case model (\ref{model}) belongs to the same universality class as
an uncorrelated random complex network with the identical $P(q)$,
e.i., degree--degree correlations does not affect the critical
behavior.

In Fig.~\ref{eigen} we present the results of numerical calculation of the
largest and the second largest eigenvalues of the branching matrix
for a scale-free degree distribution,
$P(q)=Aq^{-4}$, and uniform assortative mixing $f(q)=f>0$. One can
see that $\lambda_1$ and $\lambda_2$ are finite
even correlations are strong, $f(q)=f \gg 1$.

In order to check our predictions, Eqs.~(\ref{a1}) and (\ref{a2}), for
the amplitudes of critical and transverse modes at the percolation
threshold, we solved numerically Eq.~(\ref{xq}) for model
Eq.~(\ref{model}). We considered a simple network consisting of
two groups of vertices with degrees $q_1 =4$ and $q_2 =11$ and the following
degree distribution: $P(q_1)=0.6$, $P(q_2)=0.4$. In this situation the
branching matrix has only two modes---a critical mode and a
transverse one. The resulting dependence of the size $S$ of a giant
connected component on $p$ is shown in Fig.~\ref{fig1}. A
giant connected component emerges at $p_c=1/\lambda_1$. One can also
see a non-monotonous behavior of $dS(p)/dp$ which takes place when the
occupation probability $p$ crosses $1/\lambda_2$. Figure~\ref{fig3}
shows that the amplitudes of the critical and transverse modes near
the percolation point behave in agreement with Eqs.~(\ref{a1}) and
(\ref{a2}), i.e., $a_1 \propto (p-p_c)$ and $a_2 \propto (p-p_c)^2$.


\section{Critical behavior of strongly assortative networks}
\label{new class}

\begin{figure}[t]
\begin{center}
\scalebox{0.3}{\includegraphics[angle=0]{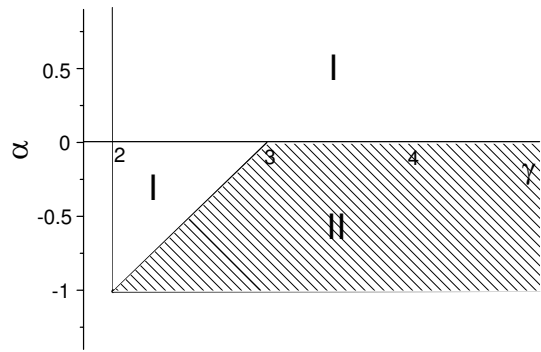}}
\end{center}
\caption{Phase diagram in the plane $(\gamma,\alpha)$ of model
(\ref{model}). Here $\gamma$ is the degree distribution exponent,
$P(q)\propto q^{-\gamma}$, and $\alpha$ is the exponent of the
function $\Lambda(q)\propto q^{-\alpha}$. In region I the
critical behavior of the model is the same as that for an
uncorrelated network with the identical degree distribution. In
region II the sequence of eigenvalues is unbounded, and this model
demonstrates a new critical exponents, see Table~\ref{table1}. The
boundary between regions I and II is defined by
the equation $\alpha = \min(0,\gamma -3)$.} \label{phase1}
\end{figure}

Let us consider the case when the function $\Lambda(q)$ is
unbounded, i.e., $\max \Lambda(q)=\infty$. Then the sequence of
eigenvalues $\lambda_i$ also is unbounded. In particular, both
$\lambda_1$ and $\lambda_2$ are infinite. The condition II is not
fulfilled. For example, we can choose the function
$\Lambda(q)=\Lambda_0 q^{-\alpha}$. It is unbound at $-1 \leq \alpha
< 0$ and bounded at $\alpha \geq 0$. In the latter case, model
(\ref{model}) has the same critical behavior as an uncorrelated
network with the identical degree distribution. In Fig.~\ref{phase1}
this region of parameters $\alpha$ and the degree distribution
exponent $\gamma$ is defined as region I.

In order to study percolation in model (\ref{model}) for $-1 \leq
\alpha < 0$, we rewrite Eq.~(\ref{xq}) in the following form:
\begin{equation}
y_q=\frac{p\Lambda(q)}{q-1}\Bigl[1-\Bigl(1-y_q
\Bigr)^{q-1}\Bigr]+\frac{z_1\rho(q)v}{q}, \label{yq2-model}
\end{equation}
where the parameter $v$ plays the role of an effective field:
\begin{eqnarray}
&&\!\!\!\!\!\! v=p \sum_{q} \rho(q)P(q) \Bigl[1-\Bigl(1-y_q
\Bigr)^{q-1}\Bigr] \nonumber
\\[5pt]
&&=\sum_{q} \frac{qy_q}{z_1 \rho(q)f(q)}\Big/\Bigl[1+ \sum_{q}
\frac{1}{f(q)} \Bigr]. \label{v-m}
\end{eqnarray}
At $p \ll 1$ and $q_{cut} \gg 1$ these equations have an approximate
solution
\begin{eqnarray}
&& y_q=\frac{p\Lambda(q)}{q-1} + \frac{z_1\rho(q)v}{q}, \,\,\,
\text{at} \,\,\, q \gg q_p, \nonumber
\\[5pt]
&&\,\,\,\,\,\,\,\, =\frac{z_1\rho(q)v}{q[1-p\Lambda(q)]}, \,\,\,
\text{at} \,\,\, q \ll q_p, \label{yq3-m}
\end{eqnarray}
where a characteristic degree $q_p$ is determined from the equation
$p\Lambda(q_p)=1$ which gives
\begin{equation}
q_p =(\Lambda_0 p)^{1/\alpha} \gg 1. \label{qp-model}
\end{equation}
Within this solution
we have $(q-1)y_q \gg
1$ at $q\gg q_p$, though $y_q \ll 1$. This enables us to use the approximation:
\begin{equation}
1-\Bigl(1-y_q
\Bigr)^{q-1} \approx 1. \label{appr}
\end{equation}
For the degree distribution $P(q)=Aq^{-\gamma}$ with
$\gamma >3$, a self-consistent solution of Eq.~(\ref{v-m}) is
\begin{equation}
v\approx \frac{Ap}{z_1 \rho (\gamma-2)q_{p}^{\gamma -2}} \propto
p^{(2 - \gamma+\alpha)/\alpha}. \label{v2-m}
\end{equation}
The leading contribution to $v$ is given by a sum over degrees $q
>q_p$. Substituting solution (\ref{yq3-m}) into
Eq.~(\ref{s}), we obtain the size of a giant connected component:
\begin{equation}
S \propto pv \propto p^{(2-\gamma+2\alpha)/\alpha}. \label{s1-model}
\end{equation}
Interestingly, the main contribution to $S$ comes from degrees $q <q_p$.

For a slowly increasing
function $\Lambda(q)=\Lambda_0 \ln  q$ (which corresponds to $\alpha \rightarrow 0$), we have $q_p=\exp
(1/(p\Lambda_0)$. Then, at an arbitrary
$\gamma >3$,
we obtain from Eq.~(\ref{v2-m}) the following results:
\begin{eqnarray}
&& v \propto p e^{-(\gamma-2)/(p\Lambda_0)}, \label{v0-model}
\\[5pt]
&& S \propto pv \propto p^2 e^{-(\gamma-2)/(p\Lambda_0)}.
\label{s0-model}
\end{eqnarray}
On the other hand, for $2< \gamma < 3$ we note
that if $q_{cut} \rightarrow \infty$, then the series
Eq.~(\ref{v-m}) is a singular function of $v$. Then,
Eq.~(\ref{v-m}) takes the form:
\begin{equation}
v\approx a p v^{\gamma-2}+\frac{Ap}{z_1 \rho (\gamma-2)q_{p}^{\gamma
-2}}, \label{v3-m}
\end{equation}
where $a=O(1)$ is a model dependent parameter. Here the first
singular term is given by the sum over $q<q_p$ under the condition
that $q_p v \ll 1$. The second term is given by degrees $q>q_p$. At
$\gamma > 3+\alpha$ we arrive at the solution Eq.~(\ref{v2-m}) while
at $2< \gamma < 3+\alpha$ we obtain
\begin{equation}
v \propto p^{\beta}
\label{v4-m}
\end{equation}
with $\beta= 1/(3-\gamma)$. The size of a giant connected component
is $S \propto pv \propto p^{1+\beta}$.
This is exactly the behavior that was found for an uncorrelated
scale-free network for $ 2< \gamma < 3$ \cite{cah02,cha03}. It means
that degree-degree correlations are irrelevant at $2< \gamma <
3+\alpha$.

Our results are summed up in Fig.~\ref{phase1} and Table
\ref{table1}. In region I, degree--degree correlations are
irrelevant. They become relevant and lead to new critical exponents
in the region of parameters $\gamma
> 3+\alpha$ and $-1 <\alpha<0 $ (region II in Fig.~\ref{phase1})
where model (\ref{model}) has an unbounded sequence of eigenvalues
of the branching matrix.

\section{Critical behavior of strongly disassortative networks}
\label{disassort}

V\'{a}zquez and Moreno \cite{Vazquez:vm03} introduced networks with
the following degree-degree distribution:
\begin{eqnarray}
&& P(q',q)=\frac{1}{z_1}(1-g_{q'})q'P(q')(1-\delta_{q',1})
\delta_{q,1} \nonumber
\\[5pt]
&& +\frac{1}{z_1}(1-g_{q})q P(q)(1-\delta_{q,1}) \delta_{q',1}
\nonumber
\\[5pt]
&&+\frac{1}{z_{1}^{2} G}g_{q'}q'P(q')g_{q}q
P(q)(1-\delta_{q',1})(1-\delta_{q,1}),
\label{2-m2}
\end{eqnarray}
where $G\equiv \sum_{q>1}g_{q}qP(q)/z_1$. The fraction of vertices
of degree 1 is fixed by the following condition:
$P(q=1)=\sum_{q>1}(1-g_{q})qP(q)/z_1$.
In this model, a vertex of degree $q>1$ is connected to a vertex of a
degree $q' =1$ with probability $1-g_q$, and this vertex ($q>1$)
is connected to a vertex of degree $q'>1$ with probability $g_q g_q'/G$.
From Eq.~(\ref{Bq}) we obtain that the average nearest-neighbor's
degree of the vertices of degree $q$ is $B(q) \propto g_q$. These
networks are disassortative for any monotonic decreasing function
$g_q$.

One can show that the branching matrix
has the largest eigenvalue
\begin{equation}
\lambda_1 =\frac{1}{z_{1}G}\sum_{q>1}g_{q}^2 q(q-1)P(q).
\label{3-m2}
\end{equation}
All other eigenvalues are zero,
$\lambda_i =0$ for $i \geq 2$.
Remarkably, this spectrum is similar to that
of an uncorrelated network, see Eq.~(\ref{unc-spectrum}). The maximal
eigenvector is
\begin{equation}
\Phi^{(1)}_q = n_1 g_q, \label{4-m2}
\end{equation}
where $n_1$ is for normalization. Using this result, we find the
local clustering coefficient $C(q)$, Eq.~(\ref{ck}), of model
(\ref{2-m2}):
\begin{equation}
C(q)=\frac{\lambda_{1}^{2}}{Nz_{1}^{2}G} q^{-2 \alpha}.
\label{5a-m2}
\end{equation}

\begin{figure}[t]
\begin{center}
\scalebox{0.3}{\includegraphics[angle=0]{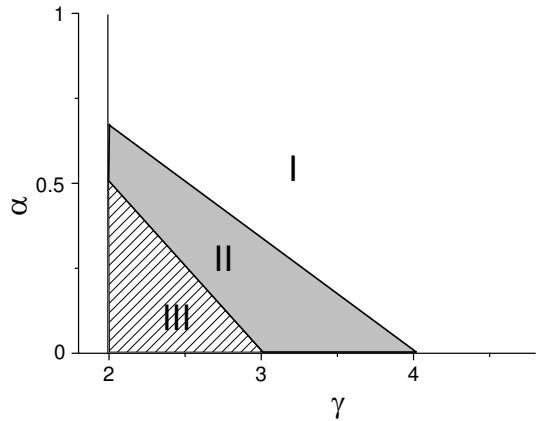}}
\end{center}
\caption{Phase diagram in the plane $(\gamma,\alpha)$ of model
(\ref{2-m2}). Here $\gamma$ is the degree distribution exponent,
$P(q)\propto q^{-\gamma}$, and $\alpha \geq 0$ is the exponent of the
function $g(q)\propto q^{-\alpha}$. In region I this model shows the standard mean-field critical singularity with the critical exponent $\beta=1$.
In
regions II  and III this model demonstrates a new critical behavior
with the critical exponent depending on $\alpha$, see
Table~\ref{table1}. The boundaries between regions I and II, and
II and III, are defined by the equations $\alpha = (4-\gamma)/3$ and
$\alpha = (3-\gamma)/2$, respectively. } \label{phase2}
\end{figure}

Let us study the case of a scale-free degree distribution $P(q)
\propto q^{-\gamma}$ and choose
\begin{equation}
g_q = q^{-\alpha} \label{5-m2}
\end{equation}
with $\alpha \geq 0$. Zero exponent $\alpha = 0$ corresponds to an
uncorrelated network. If $\alpha > 0$, then $\Phi^{(1)}_q
\rightarrow 0$ at $q \rightarrow \infty$. It means that condition III in Sec.~\ref{critical} is not fulfilled.

The dependence of the percolation threshold $p_c$ on the parameters
$\gamma$ and $\alpha$ was found in \cite{Vazquez:vm03}. Let us study
the critical behavior of model (\ref{2-m2}). From Eq.~(\ref{yq-m})
we find that the amplitudes $a_i =0$ for $i\geq2$. Therefore, the
order parameter is completely determined by the critical mode: $y_q
=a_1 \Phi^{(1)}_q$. The amplitude $a_1$ is determined by the
following equation:
\begin{equation}
(p \lambda_1 -1)a_1 =p \lambda_1 s_1(a_1), \label{6-m2}
\end{equation}
where the function $s_1(x)$ is given by series Eq.~(\ref{si-x}).
First we study the region of parameters $\alpha$ and $\gamma$, where
\begin{equation}
\alpha > \max(0,\frac{4-\gamma}{3}), \ \ \ \gamma >2
\label{7-m2}
\end{equation}
(region~I in Fig.~\ref{phase2}).
Based on Eqs.~(\ref{Mimn}) and (\ref{3-m2}) one can see that
in this region, the coefficient $M_{111}$ and
$\lambda_1$ are finite. Therefore, the percolation threshold is
non-zero, $p_c=1/\lambda_1 \neq 0$, e.i., this network is fragile
against a random damage. The  Eq.~(\ref{6-m2}) has a
solution $a_1 \propto \tau ^\beta$ with standard mean-field
exponent $\beta=1$. Remarkably, if $\alpha >2/3$, then this standard
critical behavior takes place at all $\gamma >2$. Region I
includes a subregion with $2 < \gamma < 3$ where the second moment
$\langle q^2 \rangle$ of the degree distribution diverges.
This is in contrast to an uncorrelated network
with $2 < \gamma < 3$, which has $p_c=0$ and exponent $\beta =1/(3-\gamma)$.

Region II in Fig.~\ref{phase2} is defined by the inequalities
\begin{equation}
\frac{3-\gamma}{2} < \alpha < \frac{4-\gamma}{3},  \ \ \  2< \gamma <4.  \label{8-m2}
\end{equation}
In region II, $\lambda_1$ is finite while $M_{111}=\infty$. This
points out that the function $s_1(x)$ is a singular function. The
asymptotic behavior of this function is the following: $s_{1}(x)
\propto x^{2-\nu}$ where $\nu = (4-3\alpha -\gamma)/(1-\alpha) >0$.
Equation (\ref{6-m2}) gives $a_1 \propto \tau^\beta$ with the
exponent
\begin{equation}
\beta=\frac{1 - \alpha}{\gamma - 3 +2 \alpha}. \label{9-m2}
\end{equation}

Region III in Fig.~\ref{phase2} is defined by the inequalities
\begin{equation}
0 < \alpha < \frac{3-\gamma}{2},  \,\,\, 2< \gamma <3.  \label{10-m2}
\end{equation}
In this region both $\lambda_1$ and $M_{111}$ diverge. Therefore,
$p_c=0$ and the network is robust against random damage.
To find the order parameter we directly solve Eq.~(\ref{xq}).
The exact solution is $y_q = y g_q$ where $y$
satisfy
\begin{equation}
y=\frac{p}{z_{1}G}\sum_{q>1} g_{q} q
P(q)\Bigl[1-(1-yg_q)^{q-1}\Bigr] \label{11-m2}
\end{equation}
The series in $y$ on the right-hand side
has an
asymptotic behavior $\propto y^{1-1/\beta}$, $y\ll1$, where
\begin{equation}
\beta=\frac{1 - \alpha}{3 - 2 \alpha - \gamma}.
\label{12-m2}
\end{equation}
This leads to a solution $y \propto p^{\beta}$.
The
resulting
size of a giant connected component
is
$S \propto p y$.

Thus, strong disassortative correlations can dramatically change
both the percolation threshold and the critical behavior. In a broad
region of parameters $\gamma$ and $\alpha$, see Fig.~\ref{phase2}
and Table~\ref{table1}, the critical properties of model
(\ref{2-m2}) with disassortative mixing differ from those of a
corresponding uncorrelated network.

\section{Conclusions}
\label{conclusions}

We have studied critical phenomena at the percolation transition in
equilibrium complex networks with degree-degree correlations.
Our consideration is based on the assumptions that a network is locally tree-like, i.e., clustering is negligibly small, and that only degree--degree correlations between the nearest neighboring vertices are present. The origin of degree--degree correlations is not relevant for our approach. These correlations can be 
intrinsic (i.e., they implicitly follow from other features of a network) 
or directly defined in a given network model. 

We have demonstrated that both assortative and disassortative mixing
affect not only the percolation threshold but can also change
critical behavior at this percolation point. We have found necessary
and sufficient conditions for a correlated network to have the
same critical behavior as an uncorrelated network with the identical
degree distribution. These conditions result from the fact that
critical singularities at the percolation point are determined not
only by a degree distribution, as for uncorrelated networks, but
also spectral properties of the branching matrix.  The resulting critical behavior of a correlated
network belongs to the
same universality class as percolation on an uncorrelated network
if the following conditions are fulfilled: (I) the largest eigenvalue $\lambda_1$ of the branching matrix is finite if $\langle q^2
\rangle$ is finite, or $\lambda_1 \rightarrow \infty$ if $\langle q^2 \rangle \rightarrow \infty$; (II) the second
largest eigenvalues of the branching matrix is finite; (III) the
sequence of entries of the eigenvector associated with the largest
eigenvalue converges to a non-zero value. In this situation, one can say that degree--degree
correlations (assortative or disassortative) are irrelevant for
critical phenomena though they change the value of a percolation
threshold. The critical exponents are completely determined by
asymptotic behavior of degree distribution at large degrees.

Degree--degree correlations become relevant if at least one of these
conditions is not fulfilled. We have proposed two simple models of
correlated networks with strong assortative and disassortative
mixing, where correlations dramatically change the critical
behavior. 
As a result, the critical exponent becomes model dependent and hence non-universal. The advantage of these models is that they allow us, first, to change gradually from weak 
to strong mixing, and, second, to obtain the exact analytical solution of the percolation problem. 
In the assortative networks, proposed in this work,
strong assortative mixing leads to an unbounded sequence of
eigenvalues of the branching matrix, e.i.,
condition II is not fulfilled, which results in new critical
singularities. Remarkably, in these networks the percolation
threshold can be zero despite a finite second moment $\langle q^2
\rangle$ of a degree distribution, in contrast to uncorrelated
networks. (In uncorrelated networks, $p_c =0$ only if $\langle q^2
\rangle \rightarrow \infty$.) We have also found an unusual critical
behavior in a network with strong disassortative mixing where
condition III is not fulfilled. Here the situation is
just opposite to the strongly assortative network. Namely, in
contrast to uncorrelated networks, the considered disassortative
network
can be fragile against random damage even if $\langle
q^2 \rangle \rightarrow \infty$. Moreover, we have shown that in a
wide range of model parameters, strong disassortative mixing results
in new critical singularities.

Many real-world networks and network models have 
non-zero clustering and 
long-range degree correlations. Intrinsic degree correlations are present, for example, in the static model of scale-free networks with degree exponent $2<\gamma < 3$ \cite{lgkk06} 
and in some other models without self-loops and multiple connections. 
Unfortunately, these networks are clustered, and it is unknown whether their degree-degree correlations are short- or long-ranged. A generalization of the theory to 
networks with long-range degree correlations and non-zero clustering is a challenging problem 
\cite{Dorogovtsev:dgm07}.

We believe that our general conclusions concerning a strong influence of
degree--degree correlations on critical singularities are true for
various models of statistical physics defined on the top of
correlated networks.

\begin{acknowledgments}

We thank A.~N.~Samukhin for fruitful discussions.
This work was partially supported by projects POCI: FAT/46241,
MAT/46176, FIS/61665, BIA-BCM/62662, PTDC/FIS/71551/2006, and the SOCIALNETS EU project.

\end{acknowledgments}


\appendix
\section{Spectral properties of the branching matrix}
\label{spectrum}

Here we describe general properties of the eigenvalues $\lambda_i$
and eigenvectors $\Phi_q^{(i)}$
of the branching matrix (\ref{B}). According to the
Perron-Frobenius theorem, a real positive matrix $\widehat{B}$ has a
positive real eigenvalue $\lambda_1$ such that
$\lambda_1\geq |\lambda_i|$ for any eigenvalue $\lambda_i$ of
$\widehat{B}$, see, e.g., Ref.~\cite{Minc}. Furthermore,
there is an eigenvector with positive entries $\Phi_q^{(1)}>0$
corresponding to $\lambda_1$. $\lambda_1$ is the largest
eigenvalue. The eigenvector associated with $\lambda_1$ is called a
maximal eigenvector. If this matrix is positive then the largest
eigenvalue is simple, i.e., non-generate.

One can show that a symmetric matrix
\begin{equation}
\widetilde{B}_{q,q'}= \sqrt{\frac{\langle q \rangle
(q-1)}{qP(q)}}P(q,q')\sqrt{\frac{\langle q \rangle
(q'-1)}{q'P(q')}} \label{Ts}
\end{equation}
has the same eigenvalues $\lambda_i$ as the non-symmetric matrix
$\widehat{B}$, i.e., $\sum_{q'}
\widetilde{B}_{qq'}\widetilde{\Phi}_{q'}^{(i)}= \lambda_i
\widetilde{\Phi}_q^{(i)}$, where eigenvectors
$\widetilde{\Phi}_q^{(i)}$ are related to the eigenvectors
$\Phi_q^{(i)}$ as follows:
\begin{equation}
\widetilde{\Phi}_q^{(i)}=[q(q-1)P(q)/\langle q \rangle]^{1/2}
\Phi_{q}^{(i)}. \label{2phi}
\end{equation}
This relationship between the non-symmetric matrix $\widehat{B}$ and
the symmetric matrix $\widetilde{B}$ allows us to obtain the
following spectral properties. First,
since all eigenvalues of a symmetric real matrix are real,
all eigenvalues $\lambda_i$ of $\widehat{B}$ are also real. They can be ordered as follows:
$\lambda_1\geq \lambda_2\geq ... \geq \lambda_{N_d}$. Second, the
eigenvectors $\widetilde{\Phi}^{(i)}$ form a complete orthonormal basis set:
\begin{eqnarray}
&&\sum_{q}
\widetilde{\Phi}_q^{(i)}\widetilde{\Phi}_q^{(j)}=\delta_{i,j},
\label{b1}
\\[5pt]
&& \sum_{\lambda}
\widetilde{\Phi}_q^{(\lambda)}\widetilde{\Phi}_{q'}^{(\lambda)}=\delta_{q,q'}.
\label{b2}
\end{eqnarray}
Using relation (\ref{2phi}) we obtain Eqs.~(\ref{n1}) and
(\ref{n2}). Because the maximal eigenvalue $\lambda_1$ of the
positive matrix $\widehat{B}$ is non-degenerate, this eigenvalue is
strictly larger than the second largest eigenvalue $\lambda_2$. In
other words, there is a gap between $\lambda_1$ and the second
largest eigenvalue $\lambda_2$, $\lambda_1 -\lambda_2 >0$.
A value of this gap is determined by a specific form of
a joint degree--degree distribution $P(q',q)$.

The largest eigenvalue $\lambda_1$ of a
non-negative branching matrix $B_{qq'}$ has the following upper
boundary \cite{Minc}:
\begin{equation}
\lambda_1 \leq  \max_q B(q). \label{up1}
\end{equation}
Using the equation
\begin{equation}
\lambda_1 = \max_{\Psi_q >0}
\frac{\sum_{q,q'}\Psi_{q'}(q'-1)P(q',q)(q-1)\Psi_q}{\sum_q
\Psi^{2}_{q} q(q-1)P(q)}
\label{5-lambda}
\end{equation}
which directly follows from Eq.~(\ref{sp}), we can find a lower
boundary. Substituting $\Psi_q=1$, we get
\begin{equation}
\frac{1}{z_2} \sum_q B(q)q(q-1)P(q) \leq \lambda_1. \label{lower}
\end{equation}
Unfortunately, no results are known for the second largest eigenvalue
$\lambda_2$.

 For an uncorrelated network,
$P(q',q)=q'P(q')qP(q)/\langle q \rangle^2$. In this case the
matrix $\widehat{B}$ has the entries
\begin{equation}
B_{qq'}=(q'-1)q'P(q')/\langle q \rangle \equiv B_{qq'}^{(0)}.
\label{B0}
\end{equation}
The largest eigenvalue of this matrix is
\begin{equation}
\lambda_1=z_{2}/z_1. \label{l-1-un}
\end{equation}
The normalized maximal eigenvector has equal entries
\begin{equation}
\Phi_q^{(1)}=[z_{1}/z_2]^{1/2}
\label{phi-1-un}
\end{equation}
for all $q$. There is also an ($N_d-1$)-degenerate zero
eigenvalue, $\lambda_2= ... = \lambda_{N_d}=0$. In this case the
gap, $\lambda_1 - \lambda_2$, in the spectrum is equal to
$z_{2}/z_1$. Any vector $\Phi$ orthogonal to $\Phi^{(1)}$ in the
sense of Eq.~(\ref{norma}) is an eigenvector associated with this
zero eigenvalue.

Let us consider a network in which a joint degree--degree distribution
$P(q',q)$ slightly deviates from $q'P(q')qP(q)/\langle
q \rangle^2$, i.e.,
\begin{equation}
\max_q {\frac{|B(q)-\overline{B}|}{\overline{B}}} \ll 1,
\label{weakcor}
\end{equation}
where the branching coefficient $B(q)$ and the mean branching
coefficient $\overline{B}$ are defined by Eqs.~(\ref{Bq}) and
(\ref{meanB}).

In order to find the largest eigenvalue, we write
$\widehat{B}=\widehat{B}^{(0)}+\delta \widehat{B}$, where the matrix
$\widehat{B}^{(0)}$ is given by Eq.~(\ref{B0}), and $\delta
\widehat{B}=\widehat{B}-\widehat{B}^{(0)}$ is a perturbation.
In the first order of the perturbation theory we get: 
\begin{eqnarray}
&& \Phi_q^{(1)} \approx AB(q)/\overline{B},
\label{1-phi}
\\[5pt]
&& \lambda_1 \approx
\frac{1}{z_2} \sum_q B(q)q(q-1)P(q). \label{1-lambda}
\end{eqnarray}
Here A is a normalization constant. Equation~(\ref{1-lambda}) agrees
with the lower boundary in Eq.~(\ref{lower}). Let us analyze these
results. First, the entries $\Phi_q^{(1)}$ of the maximal eigenvector are positive and lie in a bounded range.
Second, we can rewrite the right-hand side of Eq.~(\ref{1-lambda}) as
\begin{eqnarray}
&&\!\!\!\lambda_1 \approx \overline{B} + \frac{1}{\overline{B}}
\Biggl[\sum_{q,q'}(q'-1)P(q',q)(q-1)-\overline{B}^2 \Biggr]
\nonumber
\\[5pt]
&& =\overline{B} + r \sigma^2 /\overline{B}. \label{2-lambda}
\end{eqnarray}
We see that degree-degree correlations give a contribution which is
proportional to the Pearson coefficient $r$, Eq.~(\ref{r}).
Therefore, assortative mixing increases $\lambda_1$ while
disassortative mixing decreases it.
As a result we get inequality (\ref{3-lambda}).
With increasing degree--degree correlations the magnitude of the
second largest eigenvalue $\lambda_2$ increases gradually from zero.
Therefore, in a weakly correlated network the matrix $\widehat{B}$
has a finite gap $\lambda_1 -\lambda_2 >0$ which depends on a
specific degree-degree distribution $P(q,q')$.












\section{Relationship between the matrix $\widehat{B}$ and network parameters}
\label{network coefficient}

Many structural parameters of correlated networks can be related with
the spectrum of the branching matrix. The local clustering coefficient $C(q)$
of a vertex with degree $q$ is defined as follows: $C(q)\equiv
t(q)/[q(q-1)/2]$, where $t(q)$ is the number of triangles (loops of
length $3$) attached to this vertex, and $q(q-1)/2$ is the maximum
possible number of such triangles. The clustering coefficient $C$ and the mean clustering
$\overline{C}$ are given by the following equations:
\begin{eqnarray}
&&C=\frac{1}{z_2}\sum_q (q-1)P(q)C(q),
\nonumber
\\[5pt]
&&\overline{C}=\sum_q P(q) C(q). \label{c1}
\end{eqnarray}
According to Ref.~\cite{d04}, we can rewrite $C(q)$ in the following form:
\begin{equation}
C(q) =\frac{(\widehat{B}^3)_{qq}}{Nq(q-1)P(q)}.
 \label{c2}
\end{equation}
where $\widehat{B}$ is the branching matrix (\ref{B}).
Using this equation and Eq.~(\ref{n2}), we find relationships between the clustering coefficients and the eigenvalues
$\lambda_{i}$ and eigenvectors $\Phi_{q}^{(i)}$ of the branching matrix:
\begin{eqnarray}
&&C(q) = \frac{1}{N\langle q \rangle}\sum_{i=1} \lambda_{i}^3
(\Phi_{q}^{(i)})^2, \label{ck}
\\[5pt]
&&C =\frac{1}{Nz_2}\sum_{i=1}\lambda_{i}^3, \label{c}
\\[5pt]
&& \overline{C}=\frac{1}{N\langle q \rangle}\sum_{i=1} \lambda_{i}^3
\sum_q P(q) (\Phi_{q}^{(i)})^2, \label{mean-c}
\end{eqnarray}
For an uncorrelated network, $C^{(un)}=z_2 ^2/(N\langle q
\rangle ^3)$ \cite{n03}. In a network with assortative mixing the largest
eigenvalue $\lambda_1$ is larger than $z_2 /z_1$ according to
Eq.~(\ref{3-lambda}). Therefore the clustering coefficient
$C^{(as)}$ is larger than $C^{(un)}$. Similarly, in networks with
disassortative mixing the clustering coefficient $C^{(dis)}$ is
smaller then $C^{(un)}$. Thus we obtain an inequality:
\begin{equation}
C^{(dis)} < \frac{z_2 ^2}{N \langle q \rangle ^3}< C^{(as)}.
\label{ineq-c}
\end{equation}

Using Eq.~(\ref{n2}), we obtain a useful relation between the
average branching coefficient $B(q)$, Eq.~(\ref{Bq}), and the
eigenvalues $\lambda_i$ and eigenfunctions $\Phi_{q}^{(i)}$:
\begin{equation}
B(q)=v_1 \lambda_1 \Phi_{q}^{(1)} + \sum_{i\geq 2}v_i \lambda_i
\Phi_{q}^{(i)} \label{Bq-2}
\end{equation}
where $v_i =\sum_{q}w(q)\Phi_{q}^{(i)}$. This relation shows that the
degree dependence of $B(q)$ is determined by the eigenvectors
$\Phi_{q}^{(i)}$, mainly by the maximal eigenvector
$\Phi_{q}^{(1)}$. Note that if $\lambda_1\to\infty$, then $B(q\to\infty)\to\infty$ but not vice versa.

The average number, $N_{l}(q'|q)$, of vertices with degree $q'$
at a distance $l$ from a vertex of degree $q$ can also
be related with the branching matrix $\widehat{B}$:
\begin{equation}
N_{l}(q'|q)=\frac{q(\widehat{B}^l)_{qq'}}{q'-1}. \label{Nqq}
\end{equation}
The mean intervertex distance $\overline{\ell}$ of the correlated
network can be found from the equation:
\begin{eqnarray}
&&N=1+\sum_{q'}\sum_{n=1}^{\overline{\ell}}N_{l}(q'|q) \nonumber
\\[5pt]
&&=1+\langle q \rangle \sum_\lambda W(\lambda)
\sum_{n=1}^{\overline{\ell}}\lambda ^n, \label{diam}
\end{eqnarray}
where
\begin{equation}
W(\lambda)=\frac{1}{\langle q \rangle ^2}\Biggl[\sum_{q}
qP(q)\Phi_{q}^{(\lambda)}\Biggr]\Biggl[\sum_{q} q(q-1)P(q)\Phi_{q}^{(\lambda)}\Biggr].
\label{diam2}
\end{equation}
In a network with a finite gap between the largest
the second largest eigenvalues,
$\lambda_2$, the leading contribution into the right-hand side is given by the terms
with $\lambda_1$. As a result we obtain the following estimate of
$\overline{\ell}$ for a correlated network:
\begin{equation}
\overline{\ell} \approx \frac{\ln N}{\ln \lambda_1}. \label{diam2}
\end{equation}
Note that an uncorrelated network has a mean intervertex distance
$\overline{\ell} \approx \ln N / \ln (z_2 /z_1)$. Using
(\ref{3-lambda}) we find that for a given degree
distribution $P(q)$ a network with assortative mixing has a smaller
intervertex distance $\overline{\ell}^{(as)}$ than an uncorrelated
network, while a network with disassortative mixing has a larger
$\overline{\ell}^{(dis)}$:
\begin{equation}
\overline{\ell}^{(as)} < \frac{\ln N}{\ln (z_2 /z_1)}<
\overline{\ell}^{(dis)}. \label{diam2}
\end{equation}

\section{Entropy of a correlated network with a modular structure}
\label{entropy}

Here we describe a statistical ensemble of assortative networks with
degree--degree correlations, Eq.~(\ref{model}).
The probability
$P_{a}(a_{ij})$ that an edge between vertices $i$ and $j$ of degrees
$q_i$ and $q_j$ is present ($a_{ij}=1$) or absent
($a_{ij}=0$) is
\begin{eqnarray}
&&\!\!\!\!\!\!\!\!\!\! P_{a}(a_{ij})=\frac{z_{1} \rho ^2}{N}
(1+f(q_i)\delta_{q_{i},q_{j}})\delta (a_{ij}-1) \nonumber
\\[5pt]
&&\!\!\!\!\! +\Bigl[1-\frac{z_{1} \rho ^2}{N} (1+f(q_i)\delta_{q_{i},q_{j}})
\Bigr]\delta (a_{ij}). 
\label{P-a}
\end{eqnarray}
Here $a_{ij}$ are the entries of the adjacency matrix, $\rho$ and $f(q)$
are parameters. For an uncorrelated network (the configuration
model), we have $\rho=1$ and $f(q)=0$.
A positive $f(q)$ results in assortative mixing.
Let a sequence of degrees [and so a degree distribution $P(q)$] be given.
Then the probability of realization of a graph with a given adjacency matrix
$a_{ij}$ is the product of probabilities $P_{a}(a_{ij})$ over all
pairs of vertices:
\begin{equation}
{\cal P}(\{a_{ij}\})=\frac{1}{{\cal Z}}\prod_{i=1}^{N-1}%
\prod_{j=i+1}^{N}P_{a}(a_{ij})\prod_{i=1}^{N}\delta
\Bigl(\sum_{j}a_{ij}-q_{i}\Bigr). \label{config-model}
\end{equation}
The delta-functions fix degrees of the vertices. ${\cal Z}$ is a
normalization factor:
\begin{equation}
{\cal Z}=\exp \Bigl[ N \sum_{q}P(q)\ln \Bigr[\frac{1}{q!}
\Bigl(\frac{q\rho}{\rho(q)}\Bigr)^{q} \Bigr]-\frac{1}{2}Ng \Bigr].
\label{Z}
\end{equation}
where $g=\langle q \rangle [1 +\rho^2+\rho^2\sum_{q} P^{2}(q)f(q)]$.
In fact, ${\cal Z}$ is the partition function of this network
ensemble (compare with a partition function for other network
ensembles in Ref.~\cite{b07}). The function $\rho(q)$ is given by
Eq.~(\ref{rho-q}). At a given $\langle q \rangle$, parameters $\rho$
and $f(q)$ are coupled. The condition of the maximum of the entropy
$\ln {\cal Z}$ with respect to the parameter $\rho$ leads to
Eq.~(\ref{rho}).

The average of a physical quantity $A(\{a_{ij}\})$ over the network
ensemble is given by
\begin{equation}
\left\langle A\right\rangle _{\text{en}}=\int A(\{a_{ij}\}){\cal P}%
(\{a_{ij}\})\prod_{i=1}^{N-1}\prod_{j=i+1}^{N}da_{ij}. \label{en-av}
\end{equation}
In particular, for a graph with a given adjacency matrix $a_{ij}$, the degree-degree distribution
$P(q,q')$ is given by an average over edges,
\begin{equation}
P(q,q')=\frac{1}{Nz_1}\sum_{i,j}\delta_{q_{i},q} a_{ij}
\delta_{q_{j},q'}. \label{d-d-av}
\end{equation}
Averaging this function over the network ensemble, we get
Eq.~(\ref{model}).


\end{document}